\documentclass[conference]{IEEEtran}
\IEEEoverridecommandlockouts
\usepackage{cite}
\usepackage{amsmath,amssymb,amsfonts}
\usepackage{algorithmic}
\usepackage{graphicx}
\usepackage{textcomp}
\usepackage{xcolor}
\usepackage{float}
\usepackage{subfig}
\usepackage{arydshln}
\usepackage{booktabs}
\usepackage{textcase}
\usepackage{multirow}

\DeclareSubrefFormat{myparens}{#1(#2)}
\def\BibTeX{{\rm B\kern-.05em{\sc i\kern-.025em b}\kern-.08em
    T\kern-.1667em\lower.7ex\hbox{E}\kern-.125emX}}
\begin{document}

\title{An Implementation of Multimodal Fusion System for Intelligent Digital Human Generation\\

\thanks{\hrule \vspace{0.5em}
This work was supported in part by Undergraduate Training Program for Innovation and Entrepreneurship, China University of Mining and Techonology (No.202310290164Y).}

\author{
    \IEEEauthorblockN{Yingjie Zhou, Yaodong Chen, Kaiyue Bi, Lian Xiong, Hui Liu}
    \IEEEauthorblockA{School of Information and Control Engineering, China University of Mining and Technology, Xuzhou, China}
    \IEEEauthorblockA{\{zyj2000, chenyaodong, bikaiyue, xionglian, hui.liu\}@cumt.edu.cn}
}




}


\maketitle

\begin{abstract}
With the rapid development of artificial intelligence (AI), digital humans have attracted more and more attention and are expected to achieve a wide range of applications in several industries. Then, most of the existing digital humans still rely on manual modeling by designers, which is a cumbersome process and has a long development cycle. Therefore, facing the rise of digital humans, there is an urgent need for a digital human generation system combined with AI to improve development efficiency. In this paper, an implementation scheme of an intelligent digital human generation system with multimodal fusion is proposed. Specifically, text, speech and image are taken as inputs, and interactive speech is synthesized using large language model (LLM), voiceprint extraction, and text-to-speech conversion techniques. Then the input image is age-transformed and a suitable image is selected as the driving image. Then, the modification and generation of digital human video content is realized by digital human driving, novel view synthesis, and intelligent dressing techniques. Finally, we enhance the user experience through style transfer, super-resolution, and quality evaluation. Experimental results show that the system can effectively realize digital human generation. The related code is released at https://github.com/zyj-2000/CUMT\textunderscore2D\textunderscore PhotoSpeaker.\\
\end{abstract}

\begin{IEEEkeywords}
\textit{digital human, AI, deep learning, multimodality, multimedia information processing}
\end{IEEEkeywords}

\section{Introduction}
Digital humans are computer-simulated virtual images with human appearance characteristics, movement and expression, and have a wide range of application prospects in many fields such as medicine, film and virtual reality. They are considered as the entrance to the metaverse \cite{zhou2023no}. Especially in recent years, with the rapid development of computer graphics, hardware equipment and display technology, digital humans have not only become more vivid images but also more intelligent, able to help people or independently complete specific tasks. However, traditional digital human generation often includes processes such as modeling, driving, and rendering. The whole process is cumbersome and highly dependent on the designer's professional skills, aesthetic perception, and design experience. It consumes a lot of human resources and time costs. With the growing demand for digital humans, the shortcomings of traditional digital human generation technology are gradually exposed. On the other hand, thanks to the rapid development of deep learning and artificial intelligence (AI), artificial intelligence generated content (AIGC) has been able to generate text, images, and even multimedia content that is very similar to human-created content, improving the efficiency of practitioners in various industries. For this reason, it can be believed that the introduction of AI will effectively improve the efficiency of digital human generation. It is necessary and highly significant to design an intelligent digital human generation system.

\begin{figure}
    \centering
    \includegraphics[width = 8.1cm]{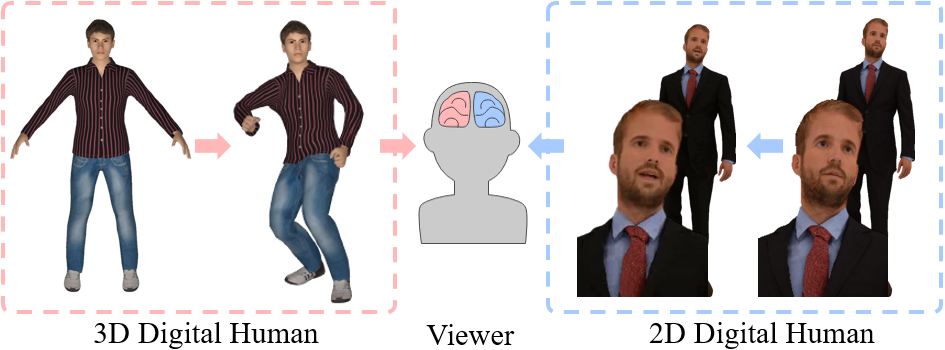}
    \caption{Classification of digital human. The selected digital humans are from the DDH-QA \cite{zhang2023ddh} and SJTU-H3D \cite{zhang2023advancing} databases.}
    \label{fig:human}
    \vspace{-0.7cm}
\end{figure}

\begin{figure*}
\includegraphics[width=\textwidth]{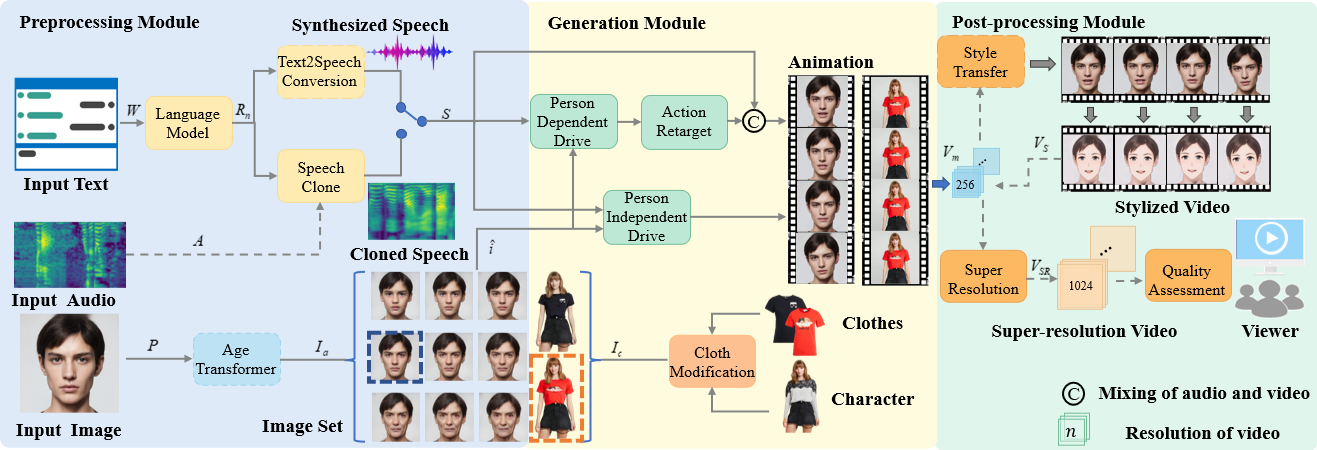}
\caption{The framework of the proposed intelligent digital human generation system.} \label{fig1}
\vspace{-0.6cm}
\end{figure*}
As shown in the Fig.~\ref{fig:human}, from the data dimension of the digital human itself, the common digital human can be divided into two-dimensional (2D) digital human and three-dimensional (3D) digital human. Between them, the 2D digital human itself is only a flat image, which usually can only be presented to the audience in the form of 2D media, such as images and videos. In addition, the research in related fields is more mature. In the last five years, many databases of 2D digital faces \cite{wang2020mead,nagrani2020voxceleb,nagrani2017voxceleb,chung2018voxceleb2,chung2016lip,afouras2018deep,son2017lip,chung2017lip} have been proposed to further promote the development of the field. Currently, there are two types of mainstream driving methods for 2D faces. Person dependent face-driving methods \cite{suwajanakorn2017synthesizing,sato2017hmm,filntisis2017photorealistic,kumar2017obamanet} are designed for specific faces, thus enabling fine-grained control with higher naturalness. Correspondingly, these advantages also lead to the limited application of person dependent methods, which are not well compatible with other faces. The person independent methods \cite{huang2023parametric,zhong2023identity,guan2023stylesync,gao2023high,li2023one,wang2023seeing,hwang2023discohead}, on the other hand, although not as effective as the person dependent methods in driving specific characters, are able to capture a wider range of patterns with more data, effectively addressing the issue of generalizability. Another type of digital human often explored in academia is the 3D digital human. A 3D digital human can be represented by point clouds, meshes or voxels, allowing the audience to view the digital human from multiple viewpoints. Thanks to the structural characteristics of 3D digital humans, they can either be rendered to generate 2D media in a specific viewpoint or placed in virtual reality (VR) for immersive experience. At present, some databases have been established in the academic field to address this issue, such as NoW \cite{sanyal2019learning}, FaceScape \cite{yang2020facescape,zhu2021facescape}, Human3.6M \cite{ionescu2013human3,ionescu2011latent}, ZJU-Mocap \cite{peng2021neural}, and BEAT \cite{liu2022beat} in the area of 3D digital humans, VOCA \cite{cudeiro2019capture} and MultiFace \cite{wuu2022multiface} in the area of speech-driven, and DHHQA \cite{zhang2023perceptual}, DDH-QA \cite{zhang2023ddh}, and SJTU-H3D \cite{zhang2023advancing} in the area of quality assessment of 3D digital humans. On this basis, the research of 3D digital humans is gradually deepened.

In this paper, we mainly focus on the 2D digital humans and use AI to provide a multimodal, interactive, 3D, stylized intelligent digital human generation technology. Specifically, the whole system contains three modules: the preprocessing module, generation module and post-processing module. The preprocessing module supports the input of rich media data consisting of text, speech and character images. After the input text is answered by language model, the speech can be obtained either through text-to-speech conversion or by cloning from the timbre of the input speech. The difference between the two methods is that the latter can preserve the phonological characteristics of the input speech. Besides, this system adopts an age transformer to customize the age for the person image, which enriches the user experience. The generation module supports person independent and person dependent digital human driving methods to modify the appearance of the digital human and generate the animation. Particularly, this system realizes the modification of digital humans' clothes. The post-processing module is mainly responsible for the style transfer and quality improvement of the generated animation. The final digital human animation can also be assessed for quality in advance through the quality evaluation model before flowing to the market. By observing the generated digital human animations, the system integrates a variety of intelligent models, which can quickly and effectively realize the generation of digital humans and meet the needs of different users. Therefore, the main contributions of this paper are as follows:

\begin{itemize}

\item An effective and feasible design method for an intelligent digital human generation system with multimodal fusion is proposed.

\item An open-source system for intelligent digital human generation with multimodal fusion realized by combining existing technologies.

\item The current status of the development of digital human-related technologies is reviewed, and application scenarios are given for this system.

\end{itemize}

\section{Proposed Method}

In this section, we specifically discuss the proposed method. The framework of the proposed method is schematically shown in Fig.~\ref{fig1}. It includes preprocessing module, generation module and post-processing module.

\subsection{Preprocessing Module}

Considering that the digital human generation system is a multimodal fusion information processing system, the preprocessing module processes the input rich media data separately for the input. Given the input media data type $M$:
\begin{equation}
M \in \{ W,A,P\} ,
\end{equation}
where $W$ denotes the input text, which is mainly used for the user to interact with the digital image, and $A$ denotes the input target audio, which is mainly used for voiceprint extraction and speech cloning. If the user does not provide the audio $A$, the system will output the digital human media with fixed voice through the text-to-speech conversion. $P$ denotes the input RGB character image, which is the main driving object. In this paper, the user input text is interactively responded to through the language model and the whole process can be described as follows: 
\begin{equation}
\begin{gathered}
  \mathcal{W}{\text{ = }}Q([{W_n},W{R_{n - 1}}, \cdots ,W{R_{1}}],U), \\ 
  {R_n}{\text{ = }}\operatorname{Re} (\mathcal{W}), \\ 
\end{gathered} 
\end{equation}
where $\mathcal{W}$ denotes a POST access request, ${W_n}$ denotes the input text of the user during the $n$th round of interaction, ${R_n}$ denotes the response text of language model during the $n$th round of interaction, ${WR_n}$ denotes the $n$th round of interaction, including a pair of input and response text, $U$ denotes the IP of the deployed language model, $Q(\cdot)$ denotes the process of a local host sending a POST access request to deployed language model, $[{W_n},W{R_{n - 1}}, \cdots ,W{R_{1}}]$ denotes the request body of the POST, and $\operatorname{Re}(\cdot)$ denotes the process where a response is fed back from the language model during the $n$th round of interaction. The response generated by the language model further selects whether or not to perform extraction of voiceprint and speech cloning based on whether or not the target audio data $A$ is input. Thus, the whole process of speech generation can be described as:
\begin{equation}
S = G_s(R_n) \parallel G_c(R_n, \mathbb{F}),
\end{equation}
where $G_s(\cdot)$ and $G_c(\cdot)$ denote the process of text-to-speech conversion and speech cloning, respectively, $\parallel$ indicates the selection of one specific content from multiple options, $\mathbb{F}$ denotes the acoustic features extracted from the target audio $A$, and $S$ is the digital human speech generated with reference to the text $R_n$. For the input digital human image $P$, the appearance of the image is selected and modified by age transformation. The whole process can be described as:
\begin{equation}
\begin{gathered}
  {I_a} = Age(P), \\ 
  {I_a} = [{i_1},{i_2},{i_3}, \cdots ,{i_k}], \\ 
\end{gathered} 
\end{equation}
where $Age(\cdot)$ denotes the process of age transformation, $I_a$ denotes the set of digital human images corresponding to each age, and $i_k$ denotes the digital human appearance image corresponding to the $k$th age.
\subsection{Generation Module}
The generation module performs clothing modification and face driving for the synthesized speech and selected digital human images, and also can widen the observation field of the digital human through the novel view synthesis of monocular RGB images to make the digital human 3D. First of all, the free dressing of the digital human can be realized through human posture detection and clothing matching. Specifically, the whole dressing process can be described as:
\begin{equation}
\begin{gathered}
  {I_c} = {C_T}(i), \\ 
  i \in {I_a}, \\ 
\end{gathered}
\end{equation}
where ${C_T}(\cdot)$ denotes the process of digital human dressing, $i$ denotes the image selected from the digital human age image set $I_a$, and $I_c$ denotes the digital human clothing image set. After that, two common driving methods are designed for the face driving of digital human. Between them, person independent driving method is simpler and can directly generate the digital human animation using the input speech and image, while person dependent driving method can only be driven for a specific person image, so it is necessary to use the action retarget after driving a specific person image in order to obtain the desired digital human animation. Overall, the whole driving process can be described as:
\begin{equation}
\begin{gathered}
  \hat i \in {I_a} \cup {I_c}, \\ 
  {V_m} = {\mathfrak{M}_i}(\hat i,S)||({\mathfrak{M}_d}(S) \oplus {\mathfrak{M}_t}(\hat i)), \\ 
\end{gathered}
\end{equation}
where, ${\mathfrak{M}_i}(\cdot)$, ${\mathfrak{M}_d}(\cdot)$ and ${\mathfrak{M}_t}(\cdot)$ denote the adoption of person independent, person dependent driving method and action retarget method, respectively, $ \oplus $ denotes the successive realization of two processes, $\hat i$ denotes the selected digital human image, $S$ is the digital human driving speech, and $V_m$ is the generated digital human animation. In particular, this system realizes the expansion of the viewpoint of the digital human through the novel view synthesis to form a 3D visual effect. This function is optional for users and the whole process can be described as:
\begin{equation}
{V_{3{\text{d}}}} = 3D({V_m}),
\end{equation}
where $3D(\cdot)$ denotes the 3D effect realized by the novel view synthesis and ${V_{3{\text{d}}}}$ denotes the synthesized 3D digital human video.

\subsection{Post-processing Module}
In order to further enhance the quality of the user experience with the digital human product, this system performs post-processing on the generated digital human animation. It is worth stating that the post-processing module is not mandatory to be used but merely provides more selectivity for users. Specifically, the module designs three parts, namely, style transfer, super-resolution, and quality assessment, respectively. Among them, the style transfer model is able to transfer the features of a specific style obtained through learning to the input media. The process can be chosen according to the actual needs and described as:
\begin{equation}
{V_s} = \Psi ({V_m}||{V_{3d}}),
\end{equation}
where $\Psi(\cdot)$ denotes the process of style transfer, and ${V_s}$ denotes the stylized video. In addition, the super-resolution technique can effectively improve the quality of digital human animation, which can be described as: 
\begin{equation}
{V_{SR}} = \varphi ({V_m}||{V_{3d}}||{V_s}),
\end{equation}
where $\varphi(\cdot)$ denotes the process of super-resolving the animation, and ${V_{SR}}$ denotes the processed video. Finally, before the digital human video is presented to the viewers, the quality assessment model may perform an objective evaluation of the generated video based on the a priori knowledge of the digital human video, which may be described as: 
\begin{equation}
Score = QA({V_m}||{V_{3d}}||{V_s}||{V_{SR}}),
\end{equation}
where $QA(\cdot)$ denotes the process of evaluating a digital human video using an objective quality assessment algorithm, and $Score$ denotes the result of the evaluation. The result can provide guidance for the improvement and optimization of the digital human generation system. 
\section{Experiments}

\begin{figure*}[tbp]
	\centering
	\subfloat[Multimodal digital human face animation]{\label{fig2:a}\includegraphics[width=7in]{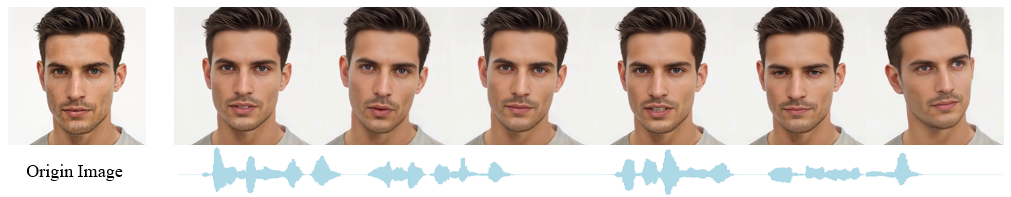}}\\
	\subfloat[Age-transformed digital human face animation]{\label{fig2:b}\includegraphics[width=7in]{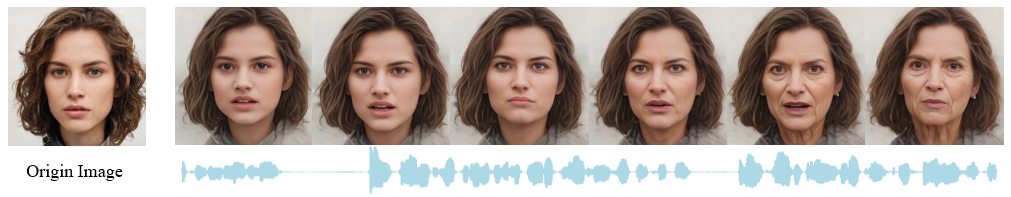}}\\
 	\subfloat[3D digital human face animation]{\label{fig2:c}\includegraphics[width=7in]{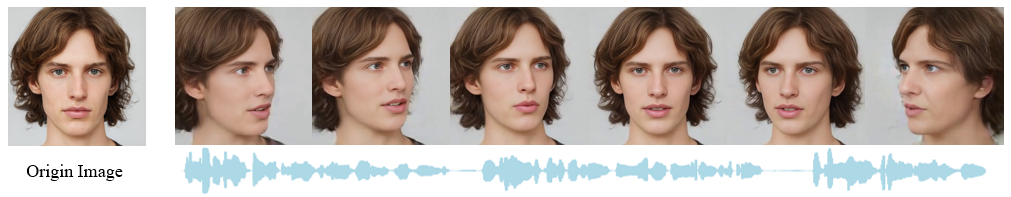}}\\
	\subfloat[Stylized digital human face animation]{\label{fig2:d}\includegraphics[width=7in]{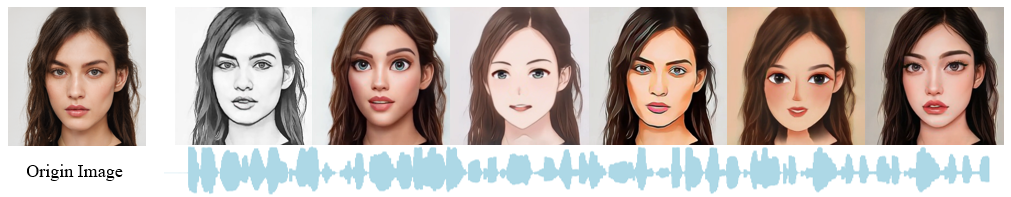}}\\	
	\caption{Visual effects of generating digital human face animations.}
        \label{vis1}
\end{figure*}
\begin{figure*}[tbp]
	\centering
	\subfloat[Multimodal digital human animation]{\label{fig3:a}\includegraphics[width=7in]{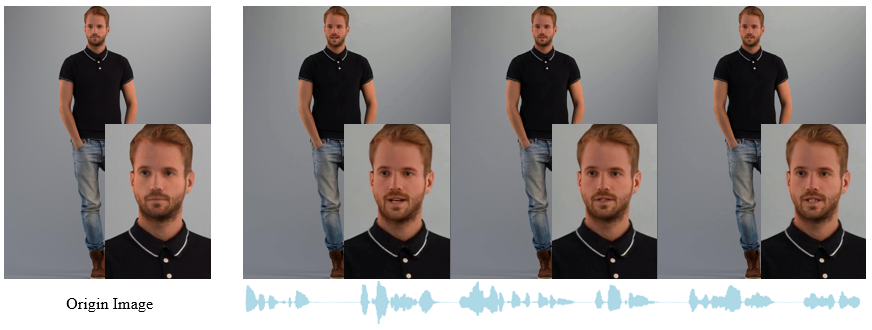}}\\
	\subfloat[Dress-up digital human animation]{\label{fig3:b}\includegraphics[width=7in]{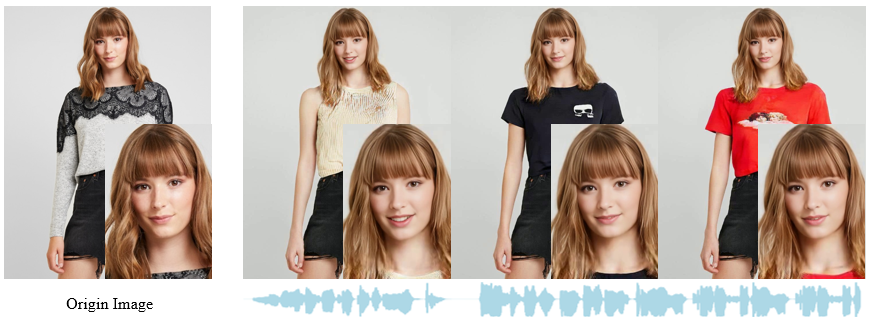}}\\
 	\subfloat[Stylized digital human animation]{\label{fig3:c}\includegraphics[width=7in]{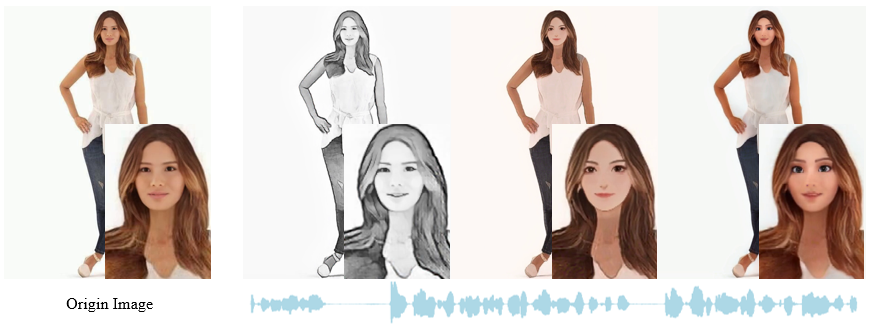}}\\	
	\caption{Visual effects of generating digital human animations.}
 \label{vis2}
\end{figure*}

\subsection{Experimental Setup}
In order to further validate the feasibility of the proposed method, we have implemented the method with the prior art, including the large language model ChatGLM \cite{zeng2022glm}, the text-to-speech conversion package eSpeaker\footnote{https://espeak.sourceforge.net/}, the speech cloning model MockingBird\footnote{https://github.com/babysor/MockingBird}, the age transformer model SAM \cite{alaluf2021matter}, the person dependent driving method LiveSpeechPortraits \cite{lu2021live}, the action retarget Thin-Plate-Spline-Motion-Model \cite{zhao2022thin}, the person independent driving method SadTalker \cite{zhang2023sadtalker}, cloth modification model VITON-HD \cite{choi2021viton}, style transfer model DCT-Net \cite{men2022dct}, super resolution model BasicVSR++ \cite{chan2022basicvsrpp,chan2022generalization}, quality assessment model VSFA \cite{li2019quality}. In our experiments, we trained and tested our proposed method on a server using Intel i7-7700K CPU @ 4.20GHz, 32GB RAM and NVIDIA 2080TI GPU.

\subsection{Generating Effects}
This section tests the visual effects of generating digital human animations, including digital face animations and digital human animations. In digital face generation, four different digital human images were selected from StyleGAN database\footnote{https://github.com/a312863063/generators-with-stylegan2}. Specifically, the generation of multimodal digital faces, age-transformed digital faces, 3D digital faces, and stylized digital faces by different audio was tested and the results are shown in Fig.~\ref{vis1}. As for digital human, we chose three digital human images from HumanAlloy\footnote{https://humanalloy.com/}, VITON-HD \cite{choi2021viton} and Cgtrader\footnote{https://www.cgtrader.com/}. Then, the generation of multimodal digital human, dress-up digital human, and stylized digital human by different audio was tested. The results are shown in Fig.~\ref{vis2}. Combining Fig.~\ref{vis1} and Fig.~\ref{vis2}, it can be concluded that the proposed method can effectively realize the generation of digital human face and full-body digital human. In addition, the method realizes the effects of age conversion, 3D, style transfer, and dress-up by prior art, which enriches the user experience.

\subsection{Performance indicators}
To further validate the generative effectiveness of the proposed method, we also tested seven generated digital human videos mentioned above using image and video quality assessment methods. In terms of image quality evaluation metrics, CPBD \cite{narvekar2011no} for evaluating blur and CGIQA \cite{zhang2023subjective} for evaluating CG animation were selected. Besides, two advanced video methods were chosen, including VSFA \cite{li2019quality} and FastVQA \cite{wu2022fast}. In this case, the raw output of CPBD was recorded, while the outputs of the CGIQA, VSFA and FAST-VQA, were all normalized. The higher the four metrics, the higher the quality of the generated digital human video. It is worth noting that the image quality assessment method evaluates each frame of the video and finally the average result is recorded. All methods are performed using source code provided by the authors. The test results are shown in Table~\ref{tab:performance}. As can be seen in Table~\ref{tab:performance}, the generated digital human videos are clear enough and have some aesthetic value. In addition, most of the generated videos achieved satisfactory performance on the two advanced video quality evaluation metrics. This further proves the effectiveness of the proposed method.

\begin{table}[t]
\renewcommand\tabcolsep{1pt}
  \caption{\MakeUppercase{Performance results for generated digital human animations.} }
  \vspace{-0.2cm}
  \begin{center}
  \begin{tabular}{c|c|cc|cc}
    \toprule
    
    \multirow{2}{*}{Index} & \multirow{2}{*}{Name} &\multicolumn{2}{c|}{Image Quality} &\multicolumn{2}{c}{Video Quality} \\ \cline{3-6}
     &  & CPBD$\uparrow$& CGIQA$\uparrow$& VSFA$\uparrow$& FAST-VQA$\uparrow$ \\  \hline
    Fig.~\subref*{fig2:a} & Short hair man face  & 0.2800& 0.6015 &0.8510 &0.8589 \\
    Fig.~\subref*{fig2:b} & Short hair woman face & 0.3391& 0.6164& 0.8880&0.8802 \\
    Fig.~\subref*{fig2:c} & Long hair man face & 0.4736&0.5622 &0.8445 &0.8694 \\
    Fig.~\subref*{fig2:d} & Long hair woman face & 0.4100&0.5688& 0.9893&0.6294 \\ \hdashline
    Fig.~\subref*{fig3:a} & Nash  & 0.4134& 0.4692& 0.8724& 0.7659\\
    Fig.~\subref*{fig3:b} & Fashion model  & 0.4628& 0.6215& 0.8647& 0.7579\\
    Fig.~\subref*{fig3:c} & Woman  & 0.4471& 0.5763& 0.8878& 0.6094\\

    \bottomrule
  \end{tabular}
  \end{center}
  \label{tab:performance}

\end{table}
\section{Conclusion}
In this paper, we propose a method for implementing an intelligent digital human generation system with multimodal fusion. The proposed method performs separate preprocessing for text, speech, and image. Then, the preprocessed media-rich data is used to generate vivid digital human videos through digital human driving methods, intelligent dressing technology, and novel view synthesis. Finally, to enhance the user experience, the post-processing module offers three optional functions: style transfer, super-resolution, and quality assessment. Furthermore, we have developed an effective and feasible multimodal fusion system for intelligent digital human generation using existing technologies based on the proposed method. The system has been validated with different images and audio, demonstrating its ability to achieve the expected results. Given the features of this system, we anticipate revolutionary applications in various fields such as entertainment, film, games, and clothing sales.

\bibliographystyle{IEEEtran}
\bibliography{IEEEexample}

\begin{thebibliography}{10}
\providecommand{\url}[1]{#1}
\csname url@samestyle\endcsname
\providecommand{\newblock}{\relax}
\providecommand{\bibinfo}[2]{#2}
\providecommand{\BIBentrySTDinterwordspacing}{\spaceskip=0pt\relax}
\providecommand{\BIBentryALTinterwordstretchfactor}{4}
\providecommand{\BIBentryALTinterwordspacing}{\spaceskip=\fontdimen2\font plus
\BIBentryALTinterwordstretchfactor\fontdimen3\font minus
  \fontdimen4\font\relax}
\providecommand{\BIBforeignlanguage}[2]{{%
\expandafter\ifx\csname l@#1\endcsname\relax
\typeout{** WARNING: IEEEtran.bst: No hyphenation pattern has been}%
\typeout{** loaded for the language `#1'. Using the pattern for}%
\typeout{** the default language instead.}%
\else
\language=\csname l@#1\endcsname
\fi
#2}}
\providecommand{\BIBdecl}{\relax}
\BIBdecl

\bibitem{zhou2023no}
Y.~Zhou, Z.~Zhang, W.~Sun, X.~Min, X.~Ma, and G.~Zhai, ``A no-reference quality
  assessment method for digital human head,'' in \emph{2023 IEEE International
  Conference on Image Processing (ICIP)}.\hskip 1em plus 0.5em minus
  0.4em\relax IEEE, 2023, pp. 36--40.

\bibitem{zhang2023ddh}
Z.~Zhang, Y.~Zhou, W.~Sun, W.~Lu, X.~Min, Y.~Wang, and G.~Zhai, ``Ddh-qa: A
  dynamic digital humans quality assessment database,'' in \emph{2023 IEEE
  International Conference on Multimedia and Expo (ICME)}.\hskip 1em plus 0.5em
  minus 0.4em\relax IEEE, 2023, pp. 2519--2524.

\bibitem{zhang2023advancing}
Z.~Zhang, W.~Sun, Y.~Zhou, H.~Wu, C.~Li, X.~Min, X.~Liu, G.~Zhai, and W.~Lin,
  ``Advancing zero-shot digital human quality assessment through text-prompted
  evaluation,'' \emph{arXiv preprint arXiv:2307.02808}, 2023.

\bibitem{wang2020mead}
K.~Wang, Q.~Wu, L.~Song, Z.~Yang, W.~Wu, C.~Qian, R.~He, Y.~Qiao, and C.~C.
  Loy, ``Mead: A large-scale audio-visual dataset for emotional talking-face
  generation,'' in \emph{European Conference on Computer Vision}.\hskip 1em
  plus 0.5em minus 0.4em\relax Springer, 2020, pp. 700--717.

\bibitem{nagrani2020voxceleb}
A.~Nagrani, J.~S. Chung, W.~Xie, and A.~Zisserman, ``Voxceleb: Large-scale
  speaker verification in the wild,'' \emph{Computer Speech \& Language},
  vol.~60, p. 101027, 2020.

\bibitem{nagrani2017voxceleb}
A.~Nagrani, J.~S. Chung, and A.~Zisserman, ``Voxceleb: a large-scale speaker
  identification dataset,'' \emph{arXiv preprint arXiv:1706.08612}, 2017.

\bibitem{chung2018voxceleb2}
J.~S. Chung, A.~Nagrani, and A.~Zisserman, ``Voxceleb2: Deep speaker
  recognition,'' \emph{arXiv preprint arXiv:1806.05622}, 2018.

\bibitem{chung2016lip}
J.~S. Chung and A.~Zisserman, ``Lip reading in the wild,'' in \emph{Computer
  Vision--ACCV 2016: 13th Asian Conference on Computer Vision, Taipei, Taiwan,
  November 20-24, 2016, Revised Selected Papers, Part II 13}.\hskip 1em plus
  0.5em minus 0.4em\relax Springer, 2017, pp. 87--103.

\bibitem{afouras2018deep}
T.~Afouras, J.~S. Chung, A.~Senior, O.~Vinyals, and A.~Zisserman, ``Deep
  audio-visual speech recognition,'' \emph{IEEE transactions on pattern
  analysis and machine intelligence}, vol.~44, no.~12, pp. 8717--8727, 2018.

\bibitem{son2017lip}
J.~Son~Chung, A.~Senior, O.~Vinyals, and A.~Zisserman, ``Lip reading sentences
  in the wild,'' in \emph{Proceedings of the IEEE conference on computer vision
  and pattern recognition}, 2017, pp. 6447--6456.

\bibitem{chung2017lip}
J.~Chung and A.~Zisserman, ``Lip reading in profile,'' in \emph{ritish Machine
  Vision Conference, 2017}.\hskip 1em plus 0.5em minus 0.4em\relax British
  Machine Vision Association and Society for Pattern Recognition, 2017.

\bibitem{suwajanakorn2017synthesizing}
S.~Suwajanakorn, S.~M. Seitz, and I.~Kemelmacher-Shlizerman, ``Synthesizing
  obama: learning lip sync from audio,'' \emph{ACM Transactions on Graphics
  (ToG)}, vol.~36, no.~4, pp. 1--13, 2017.

\bibitem{sato2017hmm}
K.~Sato, T.~Nose, A.~Ito \emph{et~al.}, ``Hmm-based photo-realistic talking
  face synthesis using facial expression parameter mapping with deep neural
  networks,'' \emph{Journal of Computer and Communications}, vol.~5, no.~10,
  p.~50, 2017.

\bibitem{filntisis2017photorealistic}
P.~P. Filntisis, A.~Katsamanis, and P.~Maragos, ``Photorealistic adaptation and
  interpolation of facial expressions using hmms and aams for audio-visual
  speech synthesis,'' in \emph{2017 IEEE International Conference on Image
  Processing (ICIP)}.\hskip 1em plus 0.5em minus 0.4em\relax IEEE, 2017, pp.
  2941--2945.

\bibitem{kumar2017obamanet}
R.~Kumar, J.~Sotelo, K.~Kumar, A.~De~Brebisson, and Y.~Bengio, ``Obamanet:
  Photo-realistic lip-sync from text,'' \emph{arXiv preprint arXiv:1801.01442},
  2017.

\bibitem{huang2023parametric}
R.~Huang, P.~Lai, Y.~Qin, and G.~Li, ``Parametric implicit face representation
  for audio-driven facial reenactment,'' in \emph{Proceedings of the IEEE/CVF
  Conference on Computer Vision and Pattern Recognition}, 2023, pp.
  12\,759--12\,768.

\bibitem{zhong2023identity}
W.~Zhong, C.~Fang, Y.~Cai, P.~Wei, G.~Zhao, L.~Lin, and G.~Li,
  ``Identity-preserving talking face generation with landmark and appearance
  priors,'' in \emph{Proceedings of the IEEE/CVF Conference on Computer Vision
  and Pattern Recognition}, 2023, pp. 9729--9738.

\bibitem{guan2023stylesync}
J.~Guan, Z.~Zhang, H.~Zhou, T.~Hu, K.~Wang, D.~He, H.~Feng, J.~Liu, E.~Ding,
  Z.~Liu \emph{et~al.}, ``Stylesync: High-fidelity generalized and personalized
  lip sync in style-based generator,'' in \emph{Proceedings of the IEEE/CVF
  Conference on Computer Vision and Pattern Recognition}, 2023, pp. 1505--1515.

\bibitem{gao2023high}
Y.~Gao, Y.~Zhou, J.~Wang, X.~Li, X.~Ming, and Y.~Lu, ``High-fidelity and freely
  controllable talking head video generation,'' in \emph{Proceedings of the
  IEEE/CVF Conference on Computer Vision and Pattern Recognition}, 2023, pp.
  5609--5619.

\bibitem{li2023one}
W.~Li, L.~Zhang, D.~Wang, B.~Zhao, Z.~Wang, M.~Chen, B.~Zhang, Z.~Wang, L.~Bo,
  and X.~Li, ``One-shot high-fidelity talking-head synthesis with deformable
  neural radiance field,'' in \emph{Proceedings of the IEEE/CVF Conference on
  Computer Vision and Pattern Recognition}, 2023, pp. 17\,969--17\,978.

\bibitem{wang2023seeing}
J.~Wang, X.~Qian, M.~Zhang, R.~T. Tan, and H.~Li, ``Seeing what you said:
  Talking face generation guided by a lip reading expert,'' in
  \emph{Proceedings of the IEEE/CVF Conference on Computer Vision and Pattern
  Recognition}, 2023, pp. 14\,653--14\,662.

\bibitem{hwang2023discohead}
G.~Hwang, S.~Hong, S.~Lee, S.~Park, and G.~Chae, ``Discohead:
  Audio-and-video-driven talking head generation by disentangled control of
  head pose and facial expressions,'' in \emph{ICASSP 2023-2023 IEEE
  International Conference on Acoustics, Speech and Signal Processing
  (ICASSP)}.\hskip 1em plus 0.5em minus 0.4em\relax IEEE, 2023, pp. 1--5.

\bibitem{sanyal2019learning}
S.~Sanyal, T.~Bolkart, H.~Feng, and M.~J. Black, ``Learning to regress 3d face
  shape and expression from an image without 3d supervision,'' in
  \emph{Proceedings of the IEEE/CVF Conference on Computer Vision and Pattern
  Recognition}, 2019, pp. 7763--7772.

\bibitem{yang2020facescape}
H.~Yang, H.~Zhu, Y.~Wang, M.~Huang, Q.~Shen, R.~Yang, and X.~Cao, ``Facescape:
  a large-scale high quality 3d face dataset and detailed riggable 3d face
  prediction,'' in \emph{Proceedings of the ieee/cvf conference on computer
  vision and pattern recognition}, 2020, pp. 601--610.

\bibitem{zhu2021facescape}
H.~Zhu, H.~Yang, L.~Guo, Y.~Zhang, Y.~Wang, M.~Huang, Q.~Shen, R.~Yang, and
  X.~Cao, ``Facescape: 3d facial dataset and benchmark for single-view 3d face
  reconstruction,'' \emph{arXiv preprint arXiv:2111.01082}, 2021.

\bibitem{ionescu2013human3}
C.~Ionescu, D.~Papava, V.~Olaru, and C.~Sminchisescu, ``Human3. 6m: Large scale
  datasets and predictive methods for 3d human sensing in natural
  environments,'' \emph{IEEE transactions on pattern analysis and machine
  intelligence}, vol.~36, no.~7, pp. 1325--1339, 2013.

\bibitem{ionescu2011latent}
C.~Ionescu, F.~Li, and C.~Sminchisescu, ``Latent structured models for human
  pose estimation,'' in \emph{2011 International Conference on Computer
  Vision}.\hskip 1em plus 0.5em minus 0.4em\relax IEEE, 2011, pp. 2220--2227.

\bibitem{peng2021neural}
S.~Peng, Y.~Zhang, Y.~Xu, Q.~Wang, Q.~Shuai, H.~Bao, and X.~Zhou, ``Neural
  body: Implicit neural representations with structured latent codes for novel
  view synthesis of dynamic humans,'' in \emph{Proceedings of the IEEE/CVF
  Conference on Computer Vision and Pattern Recognition}, 2021, pp. 9054--9063.

\bibitem{liu2022beat}
H.~Liu, Z.~Zhu, N.~Iwamoto, Y.~Peng, Z.~Li, Y.~Zhou, E.~Bozkurt, and B.~Zheng,
  ``Beat: A large-scale semantic and emotional multi-modal dataset for
  conversational gestures synthesis,'' in \emph{European Conference on Computer
  Vision}.\hskip 1em plus 0.5em minus 0.4em\relax Springer, 2022, pp. 612--630.

\bibitem{cudeiro2019capture}
D.~Cudeiro, T.~Bolkart, C.~Laidlaw, A.~Ranjan, and M.~J. Black, ``Capture,
  learning, and synthesis of 3d speaking styles,'' in \emph{Proceedings of the
  IEEE/CVF Conference on Computer Vision and Pattern Recognition}, 2019, pp.
  10\,101--10\,111.

\bibitem{wuu2022multiface}
\BIBentryALTinterwordspacing
C.-h. Wuu, N.~Zheng, S.~Ardisson, R.~Bali, D.~Belko, E.~Brockmeyer, L.~Evans,
  T.~Godisart, H.~Ha, A.~Hypes, T.~Koska, S.~Krenn, S.~Lombardi, X.~Luo,
  K.~McPhail, L.~Millerschoen, M.~Perdoch, M.~Pitts, A.~Richard, J.~Saragih,
  J.~Saragih, T.~Shiratori, T.~Simon, M.~Stewart, A.~Trimble, X.~Weng,
  D.~Whitewolf, C.~Wu, S.-I. Yu, and Y.~Sheikh, ``Multiface: A dataset for
  neural face rendering,'' in \emph{arXiv}, 2022. [Online]. Available:
  \url{https://arxiv.org/abs/2207.11243}
\BIBentrySTDinterwordspacing

\bibitem{zhang2023perceptual}
Z.~Zhang, Y.~Zhou, W.~Sun, X.~Min, Y.~Wu, and G.~Zhai, ``Perceptual quality
  assessment for digital human heads,'' in \emph{ICASSP 2023-2023 IEEE
  International Conference on Acoustics, Speech and Signal Processing
  (ICASSP)}.\hskip 1em plus 0.5em minus 0.4em\relax IEEE, 2023, pp. 1--5.

\bibitem{zeng2022glm}
A.~Zeng, X.~Liu, Z.~Du, Z.~Wang, H.~Lai, M.~Ding, Z.~Yang, Y.~Xu, W.~Zheng,
  X.~Xia \emph{et~al.}, ``Glm-130b: An open bilingual pre-trained model,''
  \emph{arXiv preprint arXiv:2210.02414}, 2022.

\bibitem{alaluf2021matter}
\BIBentryALTinterwordspacing
Y.~Alaluf, O.~Patashnik, and D.~Cohen-Or, ``Only a matter of style: Age
  transformation using a style-based regression model,'' \emph{ACM Trans.
  Graph.}, vol.~40, no.~4, 2021. [Online]. Available:
  \url{https://doi.org/10.1145/3450626.3459805}
\BIBentrySTDinterwordspacing

\bibitem{lu2021live}
Y.~Lu, J.~Chai, and X.~Cao, ``Live speech portraits: real-time photorealistic
  talking-head animation,'' \emph{ACM Transactions on Graphics (TOG)}, vol.~40,
  no.~6, pp. 1--17, 2021.

\bibitem{zhao2022thin}
J.~Zhao and H.~Zhang, ``Thin-plate spline motion model for image animation,''
  in \emph{Proceedings of the IEEE/CVF Conference on Computer Vision and
  Pattern Recognition}, 2022, pp. 3657--3666.

\bibitem{zhang2023sadtalker}
W.~Zhang, X.~Cun, X.~Wang, Y.~Zhang, X.~Shen, Y.~Guo, Y.~Shan, and F.~Wang,
  ``Sadtalker: Learning realistic 3d motion coefficients for stylized
  audio-driven single image talking face animation,'' in \emph{Proceedings of
  the IEEE/CVF Conference on Computer Vision and Pattern Recognition}, 2023,
  pp. 8652--8661.

\bibitem{choi2021viton}
S.~Choi, S.~Park, M.~Lee, and J.~Choo, ``Viton-hd: High-resolution virtual
  try-on via misalignment-aware normalization,'' in \emph{Proc. of the IEEE
  conference on computer vision and pattern recognition (CVPR)}, 2021.

\bibitem{men2022dct}
Y.~Men, Y.~Yao, M.~Cui, Z.~Lian, and X.~Xie, ``Dct-net: Domain-calibrated
  translation for portrait stylization,'' vol.~41, no.~4.\hskip 1em plus 0.5em
  minus 0.4em\relax ACM New York, NY, USA, 2022, pp. 1--9.

\bibitem{chan2022basicvsrpp}
K.~C. Chan, S.~Zhou, X.~Xu, and C.~C. Loy, ``{BasicVSR++}: Improving video
  super-resolution with enhanced propagation and alignment,'' in \emph{IEEE
  Conference on Computer Vision and Pattern Recognition}, 2022.

\bibitem{chan2022generalization}
------, ``On the generalization of {BasicVSR++} to video deblurring and
  denoising,'' \emph{arXiv preprint arXiv:2204.05308}, 2022.

\bibitem{li2019quality}
D.~Li, T.~Jiang, and M.~Jiang, ``Quality assessment of in-the-wild videos,'' in
  \emph{Proceedings of the 27th ACM International Conference on Multimedia},
  2019, pp. 2351--2359.

\bibitem{narvekar2011no}
N.~D. Narvekar and L.~J. Karam, ``A no-reference image blur metric based on the
  cumulative probability of blur detection (cpbd),'' \emph{IEEE Transactions on
  Image Processing}, vol.~20, no.~9, pp. 2678--2683, 2011.

\bibitem{zhang2023subjective}
Z.~Zhang, W.~Sun, Y.~Zhou, J.~Jia, Z.~Zhang, J.~Liu, X.~Min, and G.~Zhai,
  ``Subjective and objective quality assessment for in-the-wild computer
  graphics images,'' \emph{arXiv preprint arXiv:2303.08050}, 2023.

\bibitem{wu2022fast}
H.~Wu, C.~Chen, J.~Hou, L.~Liao, A.~Wang, W.~Sun, Q.~Yan, and W.~Lin,
  ``Fast-vqa: Efficient end-to-end video quality assessment with fragment
  sampling,'' in \emph{European Conference on Computer Vision}.\hskip 1em plus
  0.5em minus 0.4em\relax Springer, 2022, pp. 538--554.

\end{thebibliography}

\end{document}